\documentclass[journal]{IEEEtran}
\ifCLASSOPTIONcompsoc
  \usepackage[nocompress]{cite}
\else
  \usepackage{balance}
    \usepackage{cite}
\fi
\usepackage{amsthm}
\usepackage{amsmath}
\usepackage{amssymb}
\usepackage[ruled,vlined,linesnumbered]{algorithm2e}
\usepackage{array}
\usepackage{float}
\usepackage{graphicx}
\usepackage{pgfplots}
\pgfplotsset{compat=newest}
\usepackage{smartdiagram}
\usepackage{enumitem}
\newlist{steps}{enumerate}{1}
\setlist[steps, 1]{label = Step \arabic*:}
\newtheorem{theorem}{Theorem}
\newtheorem{definition}{Definition}
\begin{document}
\title{A Flexible and Lightweight Group Authentication Scheme}
\author {Yucel~Aydin,
        Gunes~Karabulut~Kurt,~\IEEEmembership{Senior Member, IEEE},\\
 	~{Enver~Ozdemir,~\IEEEmembership{Member, IEEE},} and {Halim~Yanikomeroglu,~\IEEEmembership{Fellow, IEEE}  }  %

\thanks{Y. Aydin is with Istanbul Technical University, Istanbul, 34485 TUR (email: aydinyuc@itu.edu.tr).}%

\thanks{G. Kurt is with the Department of Electronics and Communication Engineering, Istanbul Technical University, Istanbul, 34485 TUR (e-mail: gkurt@itu.edu.tr).}%

\thanks{E. Ozdemir is with the Informatics Institute, Istanbul Technical University, Istanbul, 34485 TUR (e-mail: ozdemiren@itu.edu.tr).}%

\thanks{H. Yanikomeroglu is with the Department of Systems and Computer Engineering, Carleton University, Ottawa, K1S 5B6 CAN (e-mail: halim@sce.carleton.ca).}%


\thanks{Copyright (c) 2020 IEEE. Personal use of this material is permitted. However, permission to use this material for any other purposes must be obtained from the IEEE by sending a request to pubs-permissions@ieee.org.}
} 

\maketitle
\begin{abstract}
Internet of Things (IoT) networks are becoming a part of our daily lives, as the number of IoT devices around us are surging. The authentication of millions of connected \textit{things} and the distribution and management of secret keys between these devices pose challenging research problems. Current one-to-one authentication schemes do not take the resource limitations of  IoT devices   into consideration. Nor do they address the scalability problem of massive machine type communication (mMTC) networks. Group authentication schemes (GAS), on the other hand, have emerged as novel approaches for many-to-many authentication problems. They can be used to simultaneously authenticate numerous resource-constrained devices. However, existing GAS are not energy efficient, and they do not provide enough  security for widespread use. In this paper, we propose a lightweight GAS that significantly reduces  energy consumption on devices, providing almost 80\% energy savings when compared to the state-of-the-art solutions. Our approach is also resistant to the replay and man-in-the-middle attacks. The proposed approach also includes a solution for key agreement and key distribution problems in mMTC environments. Moreover, this approach can be used in both centralized and decentralized group authentication scenarios. The proposed approach has the potential to address the fast authentication requirements of the envisioned agile 6G networks, supported through aerial networking nodes. 
\end{abstract}


\IEEEpeerreviewmaketitle

\section{Introduction}

A \textit{thing} in an Internet of Things (IoT) network can be defined as a physical or virtual node which connects to the Internet and has the ability to communicate with other nodes \cite{FGMAA,KCVA}.  
Security and privacy are crucial points in the advancement of IoT networks. The IoT paradigm extends the capabilities of the Internet to mobile and sensor networks. Each node is connected to the network and is also capable of communicating with each other. The confidentiality and integrity of data and the authentication of nodes are the main security issues for IoT networks \cite{suo}. Authenticating each node remains a critical challenge. 
Authentication is a process for ascertaining that an entity really is who it claims to be \cite{smith}.  It is one of the most important processes in the access control chain since all other security and data transmission operations follow after the authentication process. In the near future, there will be numerous connected nodes around us which are connected to us and/or to each other. Most of them will have limited computing power and battery capacity. Therefore, the use of traditional cryptographic methods will not be possible for the authentication process of resource-constrained IoT nodes. IoT networks  typically have a three-layer design, comprised of a sensing layer, a network layer, and an application layer. The sensing layer consists of IoT nodes with various sensing capabilities. The network layer aids the transmission of the sensed data to the servers. Typically, gateways are used as the devices that  provide the  connection between the application layer and sensing layer, along with routers and other packet forwarding devices. Traditional authentication methods can be used in  gateways since these devices have relatively high computational power. On the other hand, lightweight authentication solutions are needed for the sensing layer nodes. However, this problem is yet to be addressed even in 5G networks, including 3GPP Rel-16 \cite{3GPP16}.

\begin{figure}[tb]
\includegraphics[width=\linewidth]{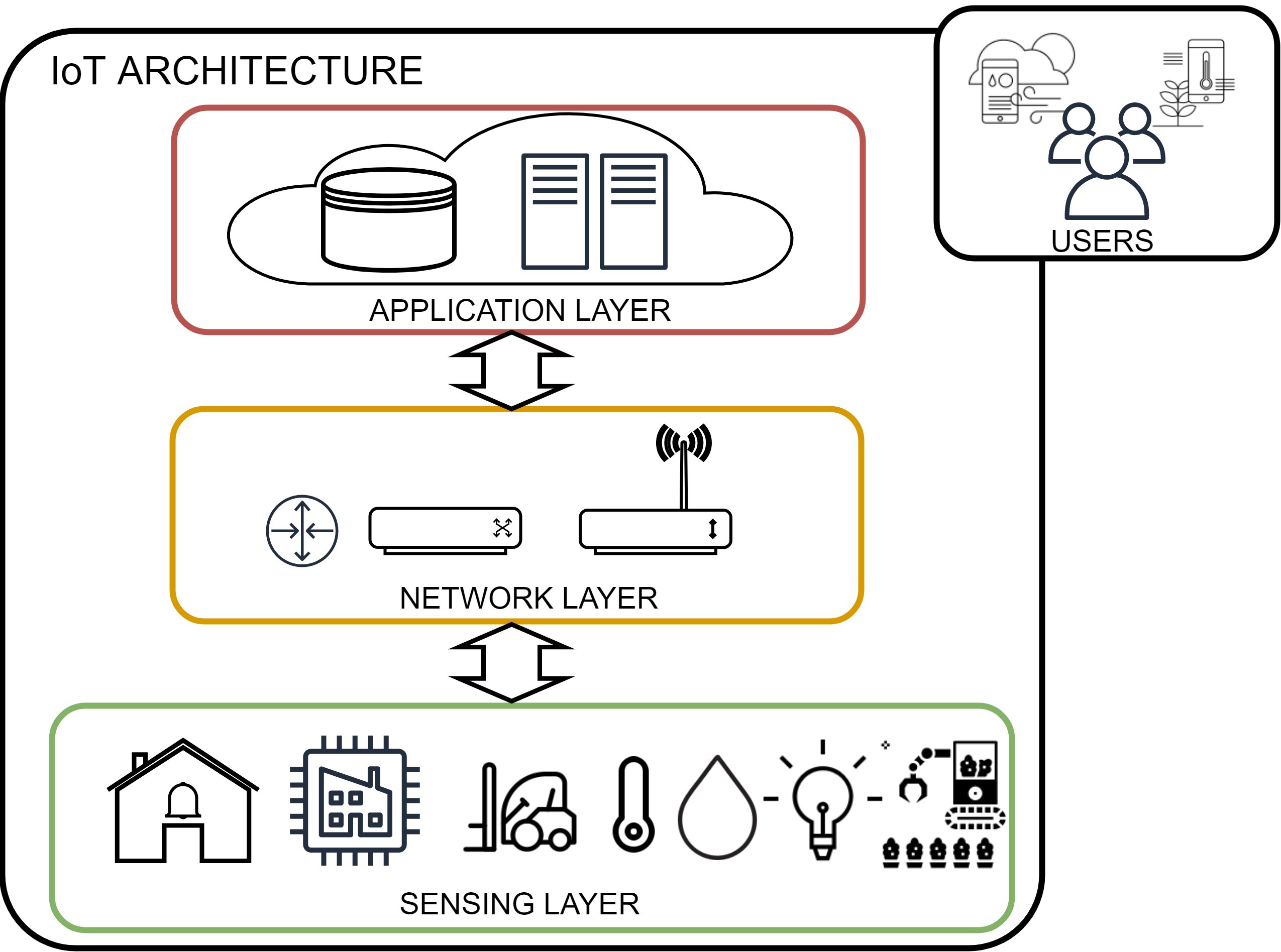}
 \caption{Three-layer Internet of Things (IoT) architecture. The data is collected through the sensing layer and transferred over the network layer to servers and databases in the application layer. Users connect to the system over the cloud.}
\label{fig:iot}
\end{figure}

 In traditional authentication methods, the client and authentication server usually have a shared key before starting the communication process. A random value, which is selected and sent by the server to the client, is encrypted by the client with the key, and the encrypted value is sent back to the server. Finally, the server validates the client by decrypting the response. In this process, there is one claimer and one prover. The prover can only authenticate one user at a time. So this approach is not scalable for densely deployed IoT networks, where millions of nodes are expected to be operational. Although the problem of providing connectivity to all these nodes is currently being addressed by 3GPP in Release 16 under the name massive machine-type communications (mMTC) the scalability of the authentication process remains to be addressed.
 
According to IMT-2020's mMTC requirements, over 1 million nodes can operate in a single km$^2$  \cite{IMT}.  Although the security issues of mMTC networks are already visible and currently being studied by the research community \cite{mtc},  to the best of the authors' knowledge, there are  no standardization efforts targeting the scalability of device authentication in these networks \cite{3GPPSec}. Each mMTC node must perform individual authentication with an authentication server according to the current evolved packet system authentication protocol (EPS-AKA) in mobile networks \cite{mtc2}. This can cause high signaling overhead on the server. 
 One of the bottlenecks in this scenario is that all IoT nodes can request authentication from the server at the same time, and the server cannot respond to all  requests. A fast authentication proposal is required to authenticate multiple users at the same time.

  Group authentication is one of the promising solutions to minimize the load on the authentication server. Millions of nodes in IoT can create groups according to their coverage area or functions in the system. Instead of sending all authentication requests to the server, authentication within the group will reduce the load of the server. The many-to-many group authentication idea was proposed by Harn in \cite{harn} and developed further by Chien in \cite{chien}.
  
  Existing  group authentication approaches do not take the  resource constraints of the network into account.  However, sensing nodes in an IoT environment frequently have limited memory, tight energy constraints, and very limited processing capability. So during the authentication process, the  communication overhead on the nodes should be as little as possible.  
 Furthermore, the energy consumption of the group authentication algorithm should be as low as possible. For this reason, traditional cryptographic systems, along with existing group authentication methods,  are not well-suited for IoT, and lightweight systems must be proposed.

Group authentication in wireless communication environments is more vulnerable to attacks by unauthorized entities. Man-in-the-middle attacks can be performed by anyone who can capture group credentials. Hence, group authentication algorithms must provide security for attacks on the wireless channel. Existing group authentication approaches remain vulnerable to such attacks.
 
Another challenge for IoT networks is the need for secure communication between nodes without any human intervention. For secure communication between millions of mMTC nodes, each node must have a private key. In such a crowded environment, key distribution and key management consume a wast amount of time and energy.

\begin{figure}[tb]
  \includegraphics[width=\linewidth]{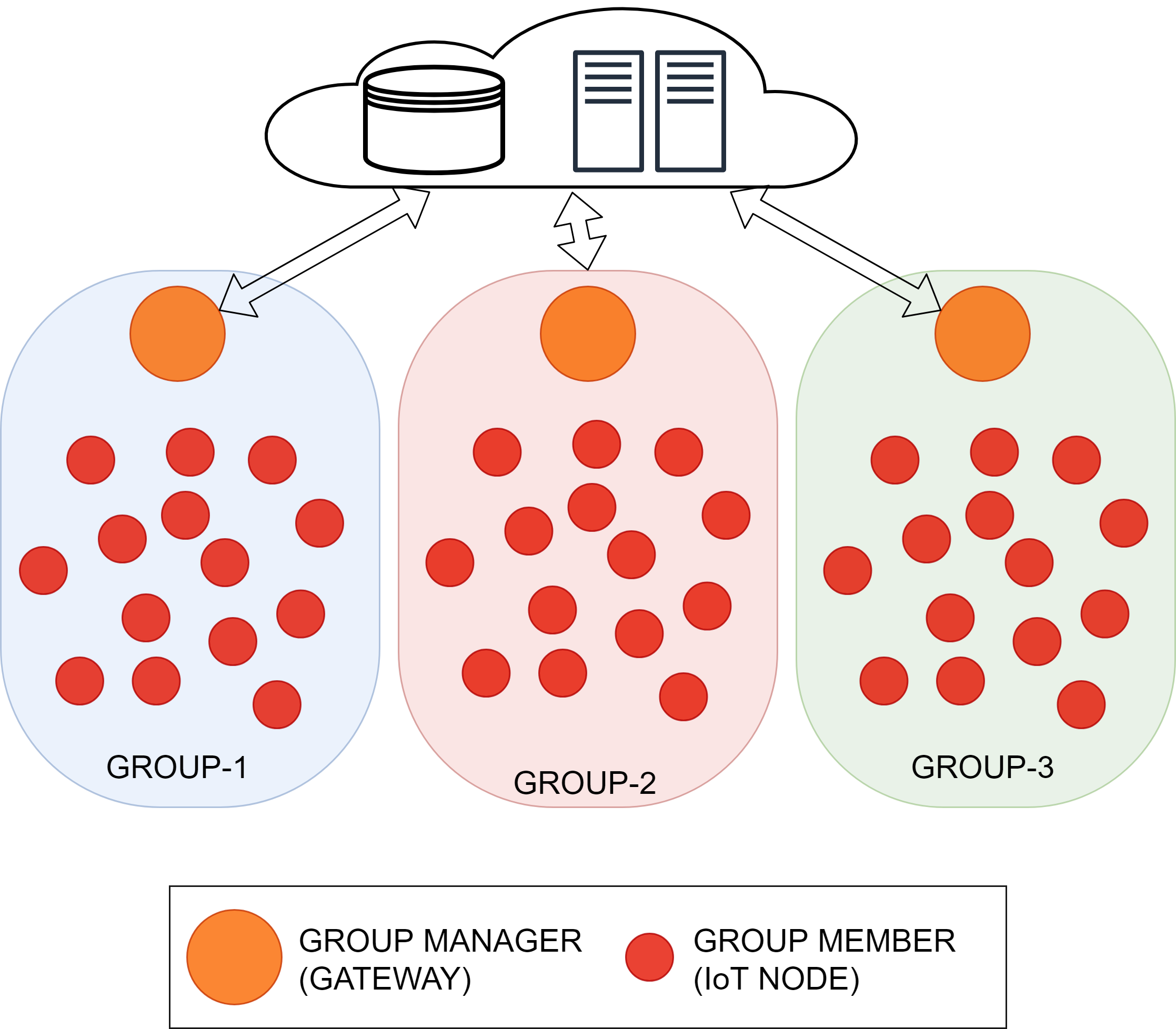}
  \caption{An exemplary  group authentication use-case. A cloud-based IoT network with different groups is shown. Each gateway acts as the GM.}
  \label{fig:usecase}
\end{figure}

  The use-case example in Figure \ref{fig:usecase} is the key motivation for our method. Cloud-based IoT technology consists of three components: the cloud network, IoT gateways, and IoT sensors. As noted, authentication is one of the most crucial steps in cloud-based IoT technology. Each component must authenticate others before transmitting any information. The cloud must authenticate the gateways, and the gateways must authenticate the sensors as well. If the number of sensors is too high in an IoT network, a group authentication scheme is vital. As shown in Figure \ref{fig:usecase}, each gateway with its sensors creates a group. The critical point here is that the gateway has a higher computational capability than sensors. The gateway should be the group manager (GM) and perform group authentication before data transmission. 

  A key agreement scheme is also required to ensure the confidentiality of the data. Below, we propose a key agreement scheme after group authentication is completed. As illustrated by the scenario, a central authority is needed for group authentication. However, if we assume that sensors will create a group without a gateway, there may not be a central authority. The appropriate selection of  the GM in this scenario becomes essential from the authentication perspective. For such an ad-hoc scenario,  we also propose group authentication in the absence of a GM. Two options are noted at each step of our proposal. Hence, we provide a flexible and lightweight group authentication scheme that also scales with the increasing number of  sensor nodes.

In light of these challenges, our main contributions are as listed:
\begin{itemize}

\item We propose a flexible lightweight authentication scheme to overcome the possible problems in group authentication and that can be used in centralized or decentralized group authentication scenarios. Group members share their private keys publicly in other group authentication schemes (GAS). If an intruder can eavesdrop on the group traffic, the intruder can recover the group key, which is the key that can be used by each group member to securely communicate with each other.

\item Our proposed approach offers a group key for mMTC. At the end of the group authentication, each group member can recover the same group key for further communications. The mMTC nodes communicate with each other by symmetric key encryption once they have a group key. In terms of privacy and security,  the private keys are not used for group authentication. This approach secures the privacy of the group members or prevents intruders from capturing private keys and performing man-in-the-middle, impersonation or replay attack. 

\item Lightweight schemes are vital for IoT scenarios due to the presence of resource constrained nodes. When we compare our proposal with other GAS, the energy consumption of one  node can be reduced by up to 80\%. Additionally, energy consumption remains constant even if the number of group members increases.

\end{itemize}

 This paper is organized as follows. The next section provides an overview of security requirements for IoT and an overview of  existing group authentication methods. In Section III, the preliminaries of our proposal are explained in detail. System and thread models are given in Section IV. Our proposed approach for group authentication is presented in Section V. The security and performance evaluation is provided in Section VI and Section VII, respectively. Conclusions are presented in Section VIII. The study is completed by an overview of future work suggestions in Section IX. 

\section{Related Works}

There are three layers in IoT architecture as shown in Figure \ref{fig:iot}. As the first layer, the sensing layer provides information from the field to the upper layers. The components in this layer include sensors of diverse applications or  radio frequency identification (RFID) tags and readers  \cite{suo}. The network layer is the mediator between the sensing and application layers and is responsible for the secure transmission of information from the former to the latter. This network
layer mostly relies on traditional networks such as the Internet, mobile communication networks, satellite networks, or wireless networks \cite{suo}. The application layer, for its part, provides services to users. Each layer may have different security requirements according to the capabilities of the layer components.  Our proposed approach focuses mostly on the sensing layer due to the computational restrictions on IoT nodes. We propose a lightweight scheme to provide authentication and key agreement for nodes in the sensing layer. Our proposed approach is based on the group authentication technique, which is used to combat scalability problems that arise as the number of components in IoT networks increases. Studies on group authentication are mostly based on secret sharing scheme, biometric solutions or aggregation of credentials, as will be detailed below.  
	 
A group of users might utilize a secret sharing algorithm for secure group communication. In this respect, a group key is divided into a number of shares via a secret sharing algorithm, and  private shares are distributed among users. Users exchange their private keys with each other, and each user can recover the group key after having its peers' private keys up to a threshold value.  Group members can communicate with each other securely by symmetric key algorithm.

The foundation of the studies in the secret-sharing area was initiated in 1979 by two different researchers. The Shamir secret sharing method was proposed by Adi Shamir in \cite{shamir}.  In the same year, the concept of key safeguarding was revealed by George Robert Blakley in \cite{blakley}. Both secret sharing and key safeguarding schemes are called threshold schemes.

Shamir's secret sharing scheme is exploited by Harn for efficient group authentication. Harn proposes three different GAS in  \cite{harn}. His first scheme is a solution if group users share their private keys with each other at the same time. Otherwise, the first scheme is not secure. The other two schemes proposed by Harn are designed for asynchronous key sharing. Chien proves in his study \cite{chien} that Harn's schemes are not secure, and an intruder can recover the security parameters. He proposes a new scheme based on elliptic curve cryptography (ECC) and bilinear mapping. Chien also compares the communication costs of Harn's approach and his proposal.

A key establishment scheme in wireless group communication is proposed in \cite{harn3}. Rather than using traditional secret sharing schemes, a linear secret sharing scheme is proposed using the Vandermonde matrix. The computational complexity of the employed secret sharing scheme is reduced in the proposed approach.

A selective group authentication scheme for IoT-based medical information systems is proposed in \cite{park}. The scheme is based on Shamir's secret sharing method. The proposal provides an authentication solution for one user to access multiple IoT nodes. There are several communications between entities in the system to authenticate only one user.

Another secret sharing method \cite{XOR} is developed using Gray code and XOR operations. The recommended method is for a group of seven users. Three or seven of the group members should come together in order to recover the master key. Although it is seen as a secure method, the authors do not explicitly state how to share the secret key securely between these members.

Asmuth and Bloom propose a key safeguarding scheme, which is based on the Chinese remainder theorem (CRT) in \cite{AB}. If anyone has shadows up to $r$, the secret value $y$ can be computed easily using CRT. But anyone with $r-1$ shadows cannot know the secret \cite{AB}.

The authors propose an algorithm using the Paillier threshold cryptography in \cite{MPP}. They compare their results with Harn's group authentication method and present experimental results. The results from \cite{MPP} show that their algorithm has a better running time result than  Harn's group authentication algorithm. However, the authors did not take into account the computational cost of public and private key encryptions or scalability issues. Note that the public key is a key that is known by everyone, but the private key is known only by the owner of the key.

 There are some studies on the usage of biometrics and machine learning techniques for group authentication. The study \cite{DLHH} proposes a shared secret knock solution for the authentication of the members in a small local group. Each group member enters a shared secret knock into the system and the system produces the training data. In the authentication phase, the group member should reproduce the same pattern of shared secret knock. It is a fast and promising solution for local groups such as family or student organizations.

Another recent study on the authentication of users by smartphones exploits biometrics \cite{WYZCW}. Both voice authentication-based and lip movement-based authentication  approaches have some vulnerabilities for the authentication of the users by smartphones. The study combines these two biometric solutions and proposes a new scheme. According to results in the paper, the considered scheme provides 95\% accuracy for user authentication and it can also detect 93\% of the attacks.

The aggregation of credentials is another solution for group authentication. The approach is mostly used for mobile networks. A trusted group member by the Long Term Evolution (LTE) network collects all the credentials from the group members. Then, the trusted member computes an aggregated value, which is the combination of the credentials. The computed value is confirmed  by the authentication server in the LTE network. Here, we also provide an overview of the research based on the aggregation of the credential approach. The authors  \cite{dgbes} propose a dynamic group based efficient and secure protocol to authenticate a group of machine type communication (MTC) devices. The protocol authenticates group members by sharing a symmetric key and verifies an aggregate message authentication code. An authentication and key agreement protocol to provide secure communication for a group of mobile station is proposed in \cite{GBAKA}. The serving network authenticates mobile stations through the home network component.

A group-based authentication protocol with dynamic policy updating for MTC in LTE networks to provide distributed authentication and session key establishment is proposed in \cite{LWZ}. The authors exploit asynchronous secret sharing and Diffie-Hellman key exchange schemes in their proposal. The GM, which has better sources than other group members, collects the credentials from the MTC devices and tries to authenticate group members through mobility management entity and home subscriber server.

The work \cite{LLZLX} proposes a lightweight group authentication scheme for machine-to-machine communication. Each MTC device computes its message authentication code and sends it to the GM. The manager authenticates the group members through the home subscriber server. A fast mutual authentication and data transfer scheme for massive narrow-band IoT devices is proposed in \cite{CYMG}. The study is also a group-based authentication scheme. A group of IoT devices can be authenticated at the same time according to the scheme.

\begin{table}[h!]
\caption{Abbreviations}
\label{table:abbreviation}
\centering
 \begin{tabular}{l l}
\hline
\textbf{Abbreviation} & \textbf{Description} \\
 \hline
IoT & Internet of Things \\
mMTC & Massive Machine Type Communication \\
GM & Group Manager \\
GAS & Group Authentication Scheme \\
CRT & Chinese Remainder Theorem \\
ECDLP & Elliptic Curve Discrete Logarithm Problem \\
ECDH & Elliptic Curve Diffie-Hellman \\
ECC & Elliptic Curve Cryptography \\
LTE & Long Term Evolution \\
UE & User Equipment \\
SSS & Shamir's Secret Sharing \\
DoS & Denial of Service \\
\hline
\end{tabular}
\end{table}

Network congestion overhead is one of the most crucial problems in the authentication of MTC devices. The authors propose an authentication and key agreement protocol for machine-to-machine communication in an IoT-enabled LTE network in \cite{PGC}. The protocol is based on the aggregation of credentials by a GM, and then group authentication is achieved between GM and authentication server.

An authentication method is proposed in \cite{BSV} for MTC to authenticate user equipments (UE) through the gateway as quickly as possible. The gateway communicates with UEs one by one to collect the tokens of UEs. The gateway authenticates UEs as a group via a home subscriber server. One-to-one communication between UEs and gateway causes additional communicational cost. 

Most of the related works are the one-to-one solution, which provides authentication for one group member at the same time. A private key is sent by a group member to the authentication server, and the server validates the private key one by one. Moreover, the related works mentioned above include energy-consuming operations.  Our proposed method is many-to-many authentication solution and also provides better energy efficiency.

{
\section{Preliminaries} 
\color{black}
This section aims to provide a brief summary of building blocks employed in the proposed algorithm. First, we present a list of abbreviations which are  used throughout the paper in Table \ref{table:abbreviation}. Considering the notation, we denote by $G$ a cyclic group, and by $K$ the characteristics of a finite field. A generator for the cyclic  group is denoted by $P$, a threshold value by $k$ or $t$, and by $f(x)$ a random function. A private key represented by $s$, and $d$ or $w$ denotes a random integer. The number of group users is denoted by $m$, the time for one multiplication in the fields $F_q$ by $T_{mul,q}$, and by $R_v$ a random point over $G$. $E$ denotes an encryption algorithm, and $D$ denotes a decryption algorithm. Hashing function is denoted by $H(\cdot)$, maximum number of users in a group by $n$, and by $ID_i$ identification number of user $i$. Bilinear pairing is denoted by $e(\cdot)$, the time for one elliptic curve multiplication by $TEM$, and by $y_t$ private key for group user $t$. 

Most of the algorithms for GAS including ours are inspired by secret sharing schemes.}
In a group $G$ of $n$ parties, a secret $D$ should be distributed in a way that at least $k<n$ parties' shares are needed to reveal the group secret $D$. This problem, in general, is solved with a secret sharing scheme or in other words a threshold scheme. A classical approximation method, which is called polynomial interpolation, is utilized for the purpose of secret sharing. The idea of the method is based on the following theorem.
\begin{theorem}\label{NThm}
	Let $(x_0,y_0),\dots,(x_k,y_k)$ be points on the graph of a function $f(x)$. There exists a unique polynomial $p(x)$ of degree $\le k$ such that $f(x_i)=p(x_i)$ for $i=0,\dots, k$ \cite{Lagrange}.
\end{theorem}  
Consider the group's secret $D$ and assume that $D$ is a real number. Theorem \ref{NThm} suggests embedding $D$ in a random polynomial $p(x)$ of degree $k-1$. In practice, the secret key $D$ is assigned to be the constant term of a polynomial $p(x)$, where other coefficients are randomly selected. Each member of the group receives a point on the graph of $p(x)$. The secret $D$ can be revealed if and only if $k$ or more members disclose their shares and this method is known as the Shamir Secret Sharing (SSS)\cite{shamir} scheme. Note that it is not feasible to recover the polynomial, even if $k-1$ points are known \cite{shamir}. In fact, in SSS, a polynomial $$p(x)=D+a_1x+...+a_{k-1}x^{k-1}$$ with degree $k-1$ is selected and first coefficient $D$ is the secret value.  $x_i$ and the corresponding $p(x_i)$ for $i=1,\dots,k$ are the shares for secret sharing scheme. Anyone who has $k-1$ distinct $(x_i,f(x_i))$ pairs cannot have knowledge about the secret, but with knowledge of $k$ or more pairs the secret value $D$ can be recovered by Lagrange interpolation formula which is $$secret=D=\sum^{t}_{i=1}p(x_i)\prod^{t}_{r=1, r\neq i}\dfrac{-x_r}{x_i-x_r}.$$
where $t\ge k$.

\begin{definition}
	Let $K$ be a field of characteristics different from 2 and 3. An elliptic curve $E$ over $K$ is an algebraic smooth curve defined by an equation of the form $y^2=x^3+Ax+B$ where $A,B\in K$.	
\end{definition}
Let $E$  be an elliptic curve over $K$. The set of all points $(x,y)$ on the curve along with a point at infinity form an abelian group $E(K)$. 
\begin{definition}
	Let $E(K)$ be an elliptic curve group for the elliptic curve $E$ over $K$. Let $P,Q$ be points in the group $E(K)$ such that $$Q=aP.$$
	For a given such points $P,Q$ the problem of finding $a$ is called discrete logarithm problem.	
\end{definition}
The use of elliptic curves in secure communication is based on the hardness assumption of the discrete logarithm problem on elliptic curve groups \cite{koblitz}. We should note here that computing in an elliptic curve group requires addition, multiplication in a field $K$ if one uses projective coordinates \cite{COH}.  \\

In our proposal, elliptic curve discrete logarithm (ECDLP) problem is frequently used.  We use ECDLP in order to provide the confidentiality of our group key $s$ via multiplying a group element $P$ with $s$ and obtaining $Q=sP$. Besides, in the group authentication phase, each user exploits ECDLP to hide its share $f(x_i)$ by multiplying a random point $P$ with $f(x_i)$ for a point $P$ on the curve.

Elliptic curve Diffie-Hellman (ECDH) key exchange protocol is also used in our proposal in the group key agreement phase. In the method, two parties want to have the same key. One party has a private integer $a$, and the other party has a private integer $b$. Their public keys are the multiplication of private integers with a public point $P$ in an elliptic curve group. Each party sends its public value to the other party in order to compute the same key. Once the first party has public key $bP$ of the second party, it computes $a$ times the public information of $bP$. The second party performs $b$ times $aP$. Finally, each party  has the same value $abP$ without releasing their secrets $a$ and $b$.

\section{Models}
\subsection{System Model}
There are $m$ group users and one GM in the group in accordance with our system model. GM has better resources than group members, such as computational power, memory, etc. GM and group members communicate with each other through wireless channels. The main problems in this system model are to authenticate all group members and to provide secure communication channels between group members. Besides, if an authentication authority tries to authenticate $m$ group members one by one, the authentication authority may collapse, and the authentication process will be time-consuming.

The structure of the groups may differ with respect to the existence of a GM. In some scenarios, the group authentication should be performed without a central authority. Therefore; group authentication scheme should be employed in both scenarios with or without a central authority.

In our proposed method, the GM begins the group authentication by selecting the initial parameters. GM selects related parameters for symmetric key encryption, elliptic curve multiplication, and ECDH key exchange protocol. In the confirmation phase, each group member computes one elliptic curve multiplication and shares a public value with all group members. If the use-case scenario is centralized, the GM performs confirmation by using the secret sharing scheme. If there is no central authority, each group member confirms the group authentication by the secret sharing scheme. After a successful group authentication, each group member shares its private keys with each other by using symmetric key encryption in order to compute a group key, which is used for the following communications.

\subsection{Threat Model}
The system model we mentioned above is vulnerable to several attacks, including man-in-the-middle, replay, denial of service attacks. A passive eavesdropper Eve may interrupt communication between one group member and GM. Then, Eve may try to authenticate herself by using the parameters obtained from a group member. She can perform man-in-the-middle and replay attacks at the same time. 

Each group member shares a computed value with GM and other group members in most of the GAS. Then, the authentication authority verifies the group members. An active attacker, Eve, may share an invalid computed value with group members and GM. Group members could not be verified due to the misused value. Eve can perform a denial of service attack by disrupting the verification phase of group authentication.

\section{The Group Authentication Framework}
Group authentication is a solution for the time and resource-consuming authentication process. In an IoT environment, the IoT nodes have limited resources, which is the reason not to handle by traditional cryptographic computations. Lightweight algorithms are widely exploited in IoT scenarios. A lightweight group authentication scheme is one of the best solutions to authenticate parties in  a communication environment with several IoT nodes.

Harn \cite{harn} and Chien \cite{chien} proposed lightweight GAS to authenticate a group of nodes at the same time. Their proposals decrease the computational load on the nodes. Our proposal addresses the shortcomings of these two most relevant studies. Below, we compare the authentication time and energy consumption in order to observe the resource consumption of IoT nodes. In the following subsections, we compare the theoretical computations of these two studies and our proposal. Then, we give the details of our proposed method step by step.
\subsection{A Comparative Overview of the Most Relevant Studies}

We have mentioned various related studies in Section II. Our proposal is nearly comparable with Harn's \cite{harn} and Chien's \cite{chien} schemes. Besides, we build our performance evaluation framework based on these two relevant studies. In this subsection, we elucidate the comparison of our work with the other two studies.

Both works are a solution for group authentication, and in general, there are one GM and multiple group users forming a group. GM determines initial parameters and keys. Each group user has one private key and one public key.

In the group authentication phase, each user computes  
\begin{equation}
c_i=\sum^{2}_{j=1}d_{i,j}f_j(x_i)\prod^{m}_{r=1, r\neq i}\dfrac{w_{i,j}-x_r}{x_i-x_r},
\end{equation}
 and  $e_i=g_i^{c_i}$ in Harn's scheme \cite{harn}. Then, each user shares $e_i$.

To verify other users, each user computes
\begin{equation}
s_i'=\prod^{m}_{i=1}e_i.
\end{equation} 
in \cite{harn}. In total, one user should make $(45m+1418)T_{mul,q}$ operations \cite{chien} ($m$ is  the number of users in the group and $T_{mul,q}$ denotes the time for one multiplication).

In Chien's scheme \cite{chien} each user computes
\begin{equation}
c_i=f(x_i)\prod^{m}_{r=1, r\neq i}\dfrac{-x_r}{x_i-x_r},
\end{equation}
and $c_i R_v$ in the group authentication phase. Then each user shares $c_i R_v$.

In the verification phase, each user computes
\begin{equation}
e\left(\sum^{m}_{i=1} c_i\times R_v,P\right){\stackrel{?}{=}}  e(R_v,Q) 
\end{equation}
in \cite{chien}.  In total, one user should make $(7m+6785)T_{mul,q}$ operations \cite{chien}.

In our proposal, each user computes $f(x_i)P$ and shares  it with other group users. The computation consumes only one elliptic curve multiplication operation, which is equal to  $(1189)T_{mul,q}$ operations \cite{chien}.

The comparison of the three studies is depicted in Table \ref{table:relevant}. As seen in the table, there are similarities between the three approaches. Both schemes have a GM to share initial parameters with group members. Besides, one group member communicates $m-1$ times with other members in all schemes. However, our scheme costs less energy for group members than others. Also, our proposed method has a solution for the de-centralized groups, which doesn't have a GM in the group.
\begin{table}[h!]
\caption{Comparison of Relevant Studies}
\label{table:relevant}
\centering
 \begin{tabular}{ ||c||c||c||c|| }
 \hline
 Parameter& Harn \cite{harn} & Chien \cite{chien} & Proposed \\
&&& Approach \\
 \hline\hline
 Presence of GM & Required & Required & Optional \\ 
&&& \\
\hline
 Total Computation &&& \\
 per user& $14m+1418$ & $7m+6785$ & $1189$ \\ 
\hline
 Total Communication &&& \\
per User & $m-1$ & $m-1$ & $m-1$ \\ 
 \hline
\end{tabular}
\end{table}

\subsection{The Proposed Method}
The GM is assumed to be infrastructure-based and has relatively more computational power. In addition to the GM, each group has several other members with the resource or computational constraints. Note that in IoT environments, the GM is basically the gateway with specific capabilities. Sensor nodes or RFID tags can be considered to be the other members of a group. The capabilities of these nodes are at a certain restricted rate. 

The proposed method has two stages. The first stage involves the authentication process, which is based on ECC and SSS. This first stage consists of two phases, which are called the initialization and confirmation phases. The second stage, which is the key agreement stage, provides a solution to construct a group key for further communications. The details of each phase  are presented below.

\begin{center}
  {\bf The Initialization Phase}
\end{center}

\begin{enumerate}
	\item GM selects a cyclic group $G$ and a generator $P$ for $G$.
	\item GM selects $E=Encryption(\cdot)$ and $D= Decryption(\cdot)$ algorithms and a hashing function $H(\cdot)$.
	\item A polynomial with degree $t-1$ is chosen by GM and the constant term is determined as group key $s$.
	\item GM selects one public key $x_i$ and one private key $f(x_i)$ for each user in the group $U$ where each user is denoted by $U_i$ for $i=1,\ldots,n$.
	\item GM computes $Q=s\times P$.
	\item GM makes $P, Q, E, D, H(s), H(\cdot), x_i$ public and shares $f(x_i)$ with only user $U_i$ for $i=1,\ldots,n$.
\end{enumerate}
    The confirmation phase is executed after the GM shares the values with the related users. There are two different options in the confirmation phase. One is that the GM is responsible for confirming the group members (\textit{the centralized approach}). The other is that any member of the group is responsible for confirming the other members (\textit{the decentralized approach}). The member selection can be made on the basis of the instantaneous resource availability of each node, such as their battery levels. 

\begin{algorithm}
{Each member computes  $f(x_i)\times P$, and shares  $f(x_i)\times P \| ID_i$ with GM and other members}\newline\\
\If{GM verifies the authentication}
{
	GM computes $f(x_i)\times P$ for each user.\newline\\
	\If{All values are valid}
	{
		Print `Authentication is complete.'
	}
	\Else{
		Repeat.
	}
}
\Else{
	{User computes $c_i$=$f(x_i)\times P{\overset{m}{\underset{r=1, r\neq i}{{\displaystyle\prod}}}(-x_r/(x_i-x_r))}$ for each user.}\newline\\
	\If{${\overset{m}{\underset{i=1}{{\displaystyle\sum}}}c_i}$ is equal to Q}
	{
		Print `Authentication is complete'.
	}
	\Else{
		Repeat.
	}
}
\BlankLine
\caption{Confirmation Phase}
\end{algorithm}

\begin{center}
{\bf The Confirmation Phase}
\end{center}
\begin{enumerate}
\item  Each user computes $f(x_i)\times P$ and sends  $f(x_i)\times P\ |ID_i$ to the GM and other users ($ID_i$ is the identification number of the user and $\|$ symbol shows the concatenation of two values). 
\item If the GM verifies the authentication, the GM computes $f(x_i)\times P$ for each user and verifies whether the values are valid or not.
\item If the GM is not included in the verification process, any user in the group computes 
\begin{equation}
C_i=\left(\prod^{m}_{r=1, r\neq i}\dfrac{-x_r}{x_i-x_r})\right)f(x_i)\times P
\end{equation}
for each user (m denotes the number of the users in the group and $m$ must be equal or larger than t). 
\item  User verifies whether
\begin{equation}
\sum_{i=1}^{m}C_i  {\stackrel{?}{=}}  Q
\end{equation}
holds.
\item If it holds, authentication is done. Otherwise, the process will be repeated from the initialization phase.
\end{enumerate}

 Both authentications by the GM and by any group member are given in Algorithm 1. It is clear that group members should only compute one elliptic curve multiplication operation. The users should send their identification numbers by concatenating with public shares in order to avoid confusion for further communications. This is because these public shares will be used by other users in further communications and in the group key agreement stage, and all members should know which public share belongs to which user.  
 
  After the authentication has been performed, users will communicate with each other by using symmetric key encryption. The key for symmetric key encryption will be calculated by senders and receivers.  

ECDH key exchange protocol is used in order to compute the key between the group members. Let us set the key, $K$ as 
\begin{equation}
K= (y_iy_jP)
\end{equation}
where $y_t=f(x_t)$, i.e., $y_t$ is the private of the user $U_t$. The sender will use their own private key $(y_i)$ and  the value sent 
by the receiver $(y_jP)$. The receiver  will obtain the same key with a similar operation, i.e., combining its own key $y_j$ with the received data $y_iP$.

  After this stage, group members can communicate with each other by symmetric key encryption method. However, using different keys for different users will cost computational and memory usage. Therefore, instead of using different keys for each user, the group key that was selected by the GM can be used as the group key. The problem is how the users will recover the group key. We basically exploit SSS and symmetric key encryption method to share the group key in the group key agreement stage.

\begin{center}
{\bf The Group Key Agreement Stage}
\end{center}
\begin{enumerate}
\item Each user shares their own private key $f(x_i)$ with other users using symmetric key encryption.
\item Each user decrypts the values and obtains $m$ different $f(x_i)$.
\item Each user computes 
\begin{equation}
s'=\sum_{i=1}^{m}f(x_i)\prod^{m}_{r=1, r\neq i}\dfrac{-x_r}{x_i-x_r}.
\end{equation}
\item Each user verifies whether 
\begin{equation}
H(s')  {\stackrel{?}{=}}  H(s)\text{ holds.}
\end{equation}
\end{enumerate}
 At the end of the group key agreement stage, each member within the group will recover the group key as given in the Algorithm 2.
  After the group key agreement process, the members of the group will be able to communicate with each other  using the group key. In addition, the GM can update $x_i$ and $f(x_i)$ values remotely using the group key in order to avoid replay attacks.  

To sum up the foregoing, we propose a comprehensive solution for the authentication of users belonging to the same group. A group authentication is accomplished with very low computational power on users in the first stage. A group key is recovered by all group users for a distributed environment in the second stage. The details of the security and performance analyses are given in the following parts. 

\begin{algorithm}
{$U_i$ computes $E_{(f(x_i)f(x_j)P)}[f(x_i)]$ for each $U_j$.}\newline\\
{Each user computes $D_{(f(x_j),f(x_i)P)}[f(x_i)]$.}\newline\\
{Each user computes 
\begin{equation}
s'=\sum_{i=1}^{m}f(x_i)\prod^{m}_{r=1, r\neq i}\dfrac{-x_r}{x_i-x_r}.
\end{equation}
}\newline\\
{Each user computes $H(s')$.}\newline\\
\If{$H(s')$ is equal to $H(s)$}
{
	Print `Group Key is recovered'.
}
\Else{
	Repeat.
}
\BlankLine 
\caption{The Group Key Agreement Stage}
\end{algorithm}

\subsection{De-centralized Scenario}
	Our proposed scheme is a solution for both centralized and decentralized scenarios. There will not always be a trusted central authority, such as GM in the distributed and crowded IoT scenarios. The IoT nodes perform the same steps in our algorithm when a GM is not present. Initial parameters can be selected in the production time of IoT nodes and they can be embedded in the nodes. The group key agreement stage is the same for both centralized and decentralized scenarios. The key phase in the de-centralized scenario is the confirmation phase. If there exists a GM in the group, the GM can confirm the credentials sent by the group members. Otherwise, if there is no GM, a multi-party computational proposal is required to establish a secure group. The de-centralized confirmation is also mentioned in Algorithm 1. After having $f(x_i)P$ public keys up to $m$, each group member can compute 
\begin{equation}
C_i=\left(\prod^{m}_{r=1, r\neq i}\dfrac{-x_r}{x_i-x_r})\right)f(x_i)\times P.
\end{equation}

	Afterward, each group member can compare 
\begin{equation}
\sum_{i=1}^{m}C_i  {\stackrel{?}{=}}  Q.
\end{equation}

	If the confirmation is done, the group members can continue with the group key agreement stage.

\section{Security Analysis}
In this section, we analyze  possible attacks to the presented algorithms above. Our proposal provides security for most man-in-the-middle and replay attacks, as shown  below.

\textbf{Theorem 1:} \textit{Group authentication cannot be performed without $t$ valid public and private values.}

\textit{Proof.} Since the stated polynomial $f(x)$ is of degree $t-1$, it is necessary to know $t$ distinct pairs   of ($x$,$f(x)$) for the formation of the polynomial again. $f(x)$ cannot be constructed  by holding less than $t$ pairs. {$ \hspace{4.1cm} \qed$}  

\textbf{Theorem 2:} \textit{The attacker who captures the value of $Q$ and $P$  sent by the GM publicly cannot have knowledge of the secret $s$.}

\textit{Proof.} Given two points $P$ and $Q$ on an elliptic curve group, it is hard to find the $s$ value that provides the relationship  $Q = s\times P$. This open problem is called ECDLP. Therefore, it is hard to find $s$ by having $Q$ and $P$. {$ \hspace{2.5cm} \qed$} 

\textbf{Theorem 3:} \textit{The attacker who captures the value of $f(x_i)P$  sent by the group members to the GM cannot have knowledge of $f(x_i)$.}

\textit{Proof.} Due to the hardness assumption of ECDLP, it is hard to find $f(x_i)$ by having $f(x_i)P$. {$ \hspace{3.4cm} \qed$} 

\textbf{Theorem 4:} \textit{The attacker can capture  $f(x_i)P$ and $f(x_j)P$ but cannot obtain a valid symmetric key in order to establish communication with user $U_i$.}

\textit{Proof.} The attacker will need $f(x_j$)  to compute $f(x_j)f(x_i)P$, but $f(x_j)$ is a secret known only by the user  $U_j$. In other words, the attacker should be able to solve computational ECDHP. {$ \hspace{4.8cm} \qed$} 
 
\textbf{Theorem 5:} \textit{The attacker cannot perform replay and man-in-the-middle attack.}
 
\textit{Proof.} An attacker can eavesdrop or cut the traffic between the GM and any user and capture $f(x_i)P$. After having $f(x_i)P$, the attacker can be part of the authentication process. Because the attacker does not have any valid $f(x_i)$, the attacker cannot communicate with any other group member by using symmetric key encryption after the authentication process. 
{$ \hspace{7cm} \qed$} 
 
Below we also list the vulnerabilities associated with the proposed approach. 
 
\textbf{Vulnerability 1:} \textit{The attacker can perform a denial of service (DoS) attacks for the authentication process.}
 
\textit{Proof.} An attacker can share a non-valid value when the members send their shares to the GM, who will confirm the authentication. GM cannot compute the group secret value and repeat the process. An attacker can share the non-valid value again and perform a denial of authentication. {$ \hspace{1.6cm} \qed$} 

\textbf{Vulnerability 2:} \textit{The node compromise attack can be performed.}
 
\textit{Proof.} If the attacker could physically capture a group member, then the attacker might obtain the private key of the member. As a result of capturing  the private key, the attacker can generate a valid public key and share it with the  GM in order to authenticate itself. If the attacker has a private key, the attacker can also communicate with the other members of the group by producing symmetric keys. {$ \hspace{2.8cm} \qed$} 
 
\textbf{Vulnerability 3:} \textit{The group members can perform DoS attacks for confirmation point.}
 
\textit{Proof.} If the group members send their shares to  the  confirmation point at the same time, this certain point can be locked. The solution for this kind of DoS attack is still a challenge in group authentication research.
{$ \hspace{3.8cm} \qed$} 

The joint design of group authentication with an authentication handoff between different groups is also crucial for an environment with millions of nodes. The groups exchange several nodes between each other. Repeating the group authentication process for each new node is resource and time-consuming. One of our future works will cover the design of a handoff scheme in mMTC or cellular environments, as noted in Section VII.
    
\section{Performance Analysis}
Group authentication is a novel method to increase the performance of the authentication system and to decrease the computational load on the group members. Additionally, the number of communications between GM and group members is kept to a minimum in group authentication. We explained the theoretical background of our simulation before giving the details of the results in the next subsection.

\subsection{Theoretical Background}
The comparison of Harn's and Chien's GAS is given in \cite{chien}. Chien used a theoretical approach for the comparison. The author unveiled the required time to complete the group authentication for both studies. $T_{mul,q}$ value, (which is the time for one multiplication in the fields $F_q$ where $q$ is 160 bits), is used as the base factor. According to \cite{chien}, $(7m+6785)T_{mul,q}$ is required to complete Chien's algorithm and $(45m+1418)T_{mul,q}$ is required to complete Harn's algorithm ($m$ is the number of users in the group).

In our proposed approach, the group members should only compute one elliptic curve point multiplication ($TEM$). According to Chien \cite{chien}, $TEM$ is roughly equal to $29T_{mul,p}$ ($T_{mul,p}$ denotes the time for one multiplication in field $p$ where $p$ is 1024 bits). The security of ECC with a 160-bit key
is roughly equal to that of RSA with a 1024-bit key or DH algorithm with a 1024-bit key \cite{ECCandRSA}. Therefore, $T_{mul,p}$ is roughly equal to 41$T_{mul,q}$ \cite{chien}. In our authentication algorithm, group members compute 29$T_{mul,p}$, which is 1189 (29x41)$T_{mul,q}$. Due to this theoretical analysis, the simulation results are shown that our scheme costs a shorter authentication time and consumes less energy than the approaches proposed by Harn and Chien.

\subsection{Results}
\begin{figure}[tb]
  \includegraphics[width=\linewidth]{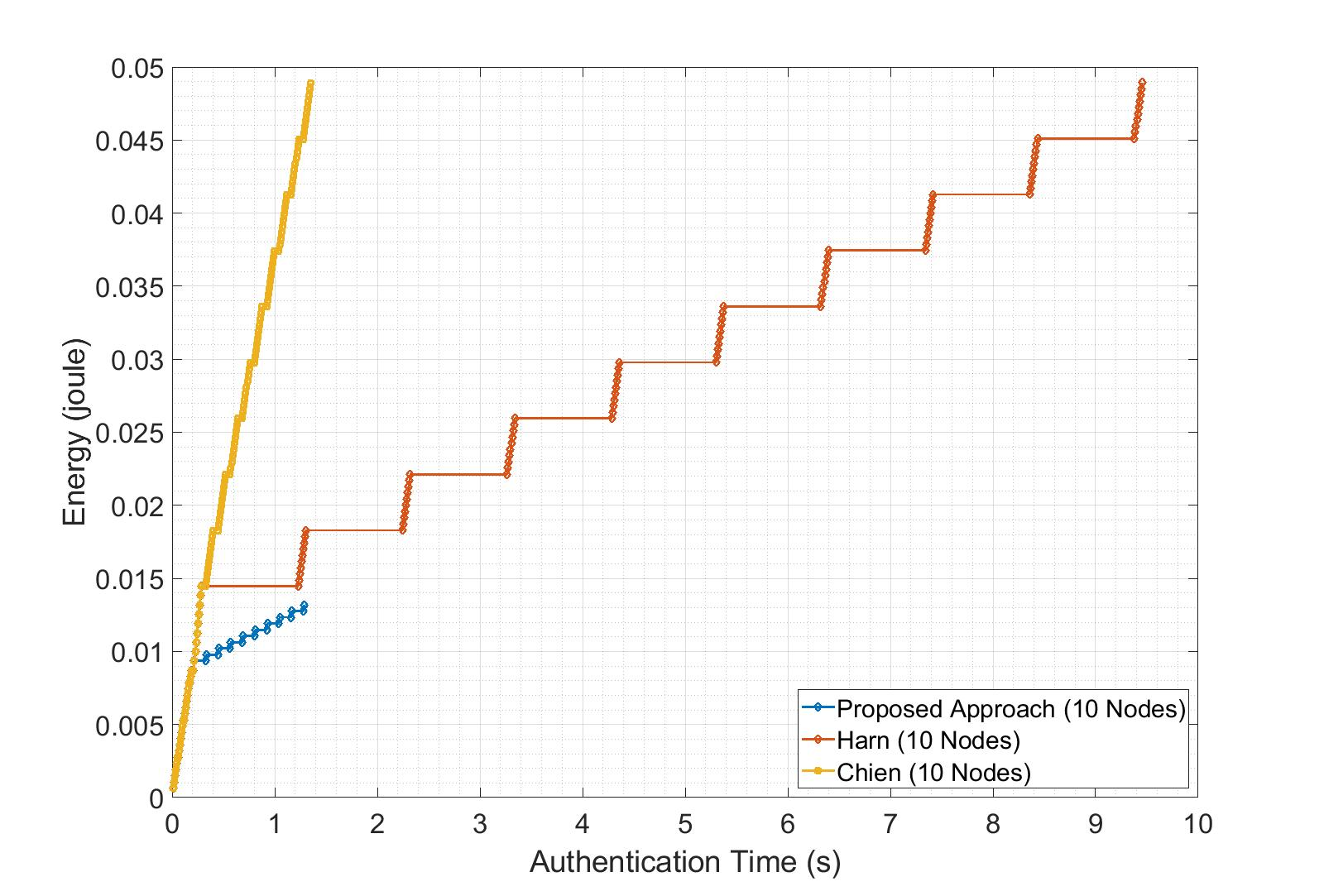}
  \caption{Energy Consumption of one IoT Node in a group with 10 IoT Nodes.}
  \label{fig:sensorConsumption10}
\end{figure}

\begin{figure}[tb]
  \includegraphics[width=\linewidth]{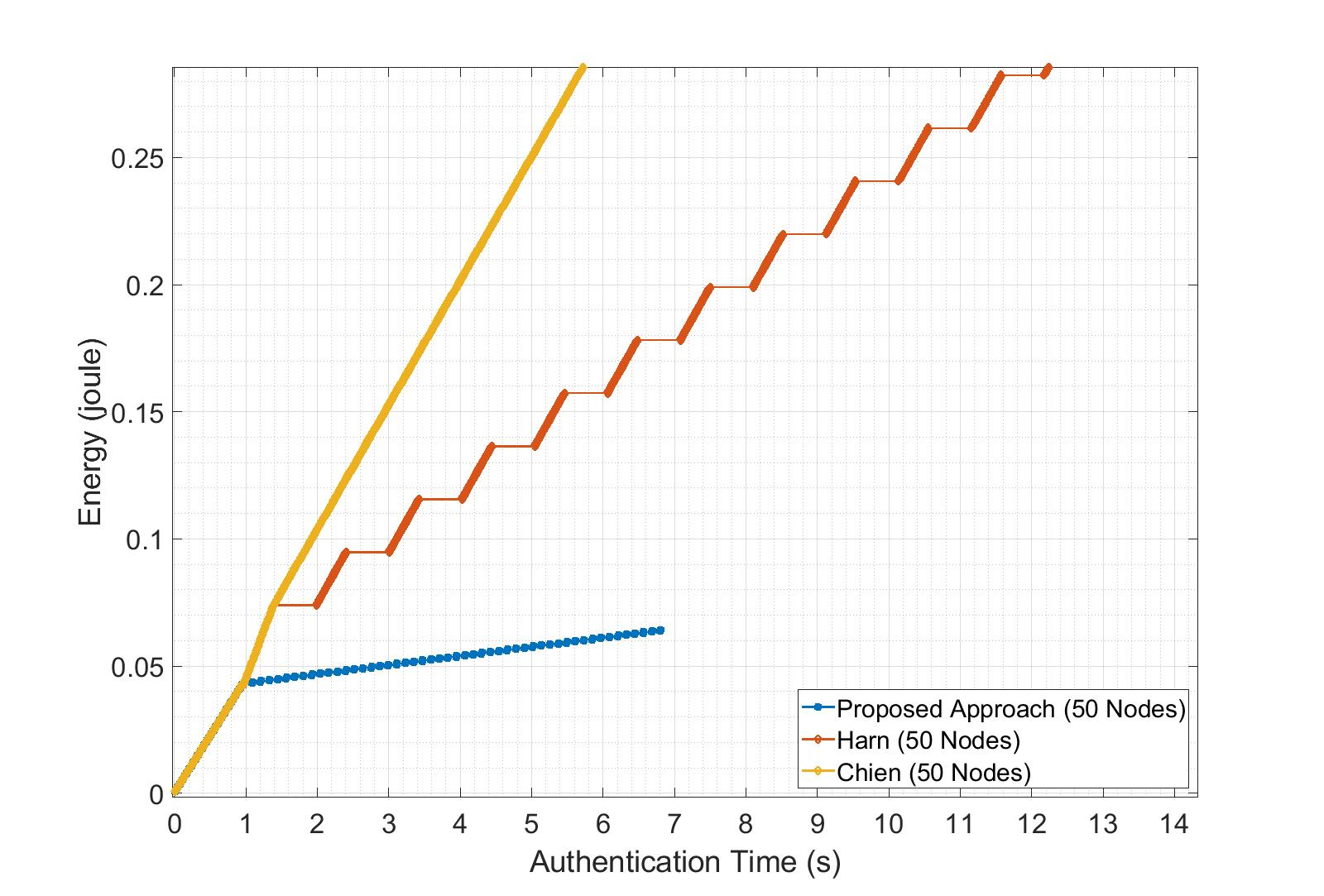}
  \caption{Energy Consumption of one IoT Node in a group with 50 IoT Nodes.}
  \label{fig:sensorConsumption50}
\end{figure}

We implemented two most relevant schemes \cite{harn,chien} and our scheme  in order to compare the energy consumption by IoT nodes. Omnet++ simulation environment \cite{omnetpp}, which is widely used to simulate wireless schemes, was exploited for the implementation of algorithms. The simulation results are given in Figure \ref{fig:sensorConsumption10} and Figure \ref{fig:sensorConsumption50} for the groups with ten IoT nodes and fifty IoT nodes.

Initial parameters of simulation were selected according to the basics in the related papers.  For Harn's scheme, the prime numbers $p$ and $q$ are two primes such that $p-1=2q$. $w_i$ and $d_i$ values are random integers that are used for each user and each secret calculation. Generator $g_i$ of field $q$  is $7$, and coefficients of two polynomials are in the field of $q$.

For Chien's and our proposal simulation, the same parameters were selected. Elliptic curve is $y^2=x^3+6x+36$ mod $2017$ selected in order to have fast computation. {\color {black} Coefficients of polynomial $f(x)$ are} in the field of $q$.

Omnet++ simulation application offers various different configurations, and the configuration we used for our simulation can be seen in Table \ref{table:channel}. IoT nodes were selected sensor node as in the omnetpp inet library. Sensors use the default options for energy storage and consumption.
\begin{table}[h!]
\caption{Simulation Parameters}
\label{table:channel}
\centering
 \begin{tabular}{ ||c||c|| }
 \hline \hline
Radio Medium & UnitDiskRadioMedium \\
\hline
IoT Node Range & 500m \\
\hline
Wlan Type & AckingWirelessInterface \\
\hline
Energy Storage Type & IdealEpEnergyStorage \\
\hline
Energy Consumer Type & SensorStateBasedEpEnergyConsumer \\
\hline
Wlan Mac Type & CsmaCaMac \\
\hline
BitRate & 1 MBPS \\
\hline
Ack Usage & False \\
\hline\hline
\end{tabular}
\end{table}

It can easily be observed from the graphics that Harn's scheme takes more time than other schemes to complete the group authentication. Time is directly proportional to the number of IoT nodes. Our proposal and Chien's scheme are almost consuming the same amount of time, which is 1.3 seconds for ten nodes and 6.9 seconds for fifty nodes. Harn's scheme consumes 10 seconds for ten nodes and 50 seconds for fifty nodes, as seen in Table \ref{table:authenticationTime}.

In terms of energy consumption, our scheme costs the least energy both for the groups with ten nodes and fifty nodes. Harn's and Chien's schemes consume almost the same energy if the group is with ten nodes. Our scheme consumes 0.014 joules of energy, whereas other schemes consume 0.05 joules. If the number of group nodes increases,  Harn's scheme consumes the most energy to complete group authentication. For groups with fifty nodes, Chien's scheme consumes 0.37 joules and  Harn's scheme consumes 1.1 joules. The nodes consume only 0.062 joules in our proposal for the group authentication if the group is with fifty nodes.
\begin{table}[h!]
\caption{Authentication Time}
\label{table:authenticationTime}
\centering
 \begin{tabular}{ ||c||c||c||c|| }
 \hline
 Time (s) & Harn \cite{harn} & Chien \cite{chien} & Proposed \\
&&& Approach \\
 \hline\hline
 10 Nodes & 10 seconds & 1.3 seconds & 1.3 seconds \\ 
\hline
50 Nodes & 50 seconds & 6.9 seconds & 6.9 seconds \\
 \hline
\end{tabular}
\end{table}

In respect to algorithmic details, IoT nodes exploit modulo exponential operations for group authentication in Harn's algorithm. The operations consume too much time and energy, which can be observed in Figure \ref{fig:sensorConsumption10} and Figure \ref{fig:sensorConsumption50}. If the number of IoT nodes increases, one IoT node consumes more time and energy. Chien's algorithm yields better results than Harn's algorithm. This is observed since the nodes only compute one elliptic curve multiplication operation and $m-1$ modulo multiplication and inverse operations ($m$ is the number of group nodes). The difference between our algorithm and Chien's algorithm is that IoT nodes compute only one elliptic curve multiplication operation for group authentication. As seen in Figure \ref{fig:sensorConsumption10} and Figure \ref{fig:sensorConsumption50}, our proposal consumes less time and energy than Chien's scheme.

\section{Conclusion}
 This study proposed a novel method for authentication of group communication in mMTC networks. Many-to-many authentication is used for group authentication by several
studies, but resource-constrained nodes were forced to compute
more than their capacity. In the proposed method, group members should only compute
one elliptic curve point multiplication. The secret sharing scheme and ECC are used on the basis of the proposed algorithms.  Most of the resource-consuming work is done by the GM or one of the group members, not by all the group members.  

Both centralized and decentralized scenarios are supported. 
In the centralized group authentication method, most of the resource-consuming work is assigned to the GM. Thus, energy-constrained group members do not need to consume a high amount of energy. In the decentralized group authentication method, only one of the group members, possibly with more computational capacity or more battery capacity than others, consumes more energy. The rest of the group members have  a limited computational load, enabling them to function in an energy-efficient manner. 

In the existing group authentication approaches, attack possibilities still exist. To the best of the authors' knowledge, there are no proposals against replay, node compromise or DoS attacks under the framework of secret sharing schemes. Our proposal provides security against replay attacks if the GMs update the credentials for each authentication. 

\section{Open Research Problems}

Each group in the group authentication scheme is coordinated by the GM. This architecture resembles the cellular model, where the communication within each cell is monitored and enabled through the base station. As nodes can be mobile in some IoT networks, defining a handoff process for mobile devices to be authenticated enables fast and reliable communication, possibly with low latency. Hence, the joint design of group authentication with the authentication handoff is of critical importance to accommodate a large number of users that may also be mobile. Instead of repeating the group authentication process from the start, the new node must be authenticated by  using a fast handoff scheme. For future works, such handoff schemes must be designed to support mobility in mMTC environments. 

Another research problem lies in the configuration of the base stations. Current networks make use of  fixed base stations to enable authenticated communication between mobile users. However, in beyond-5G wireless networks, an ultra-agile radio access architecture with mobile base stations (including aerial base stations) is envisioned as a solution for coverage and/or unexpected congestion problems \cite{bor}.  One of the deployment challenges for such dynamic cell structures will be from the authentication perspective. Although current networks do not need the authentication of the base station,  in the presence of a mobile aerial base station, these intermediate devices also need to be authenticated to avoid man-in-the-middle attacks.  A fast authentication scheme will be required to provide authentication between mobile ground users and aerial base stations and/or between the aerial base station and the terrestrial base stations. Moreover, mutual authentication of all active  aerial base stations will also be necessary. Furthermore, the handoff processes of all the aforementioned connections will need to be carefully designed.  We believe that the proposed group authentication technique can serve as a foundation for these open research problems.  

\balance

\begin{IEEEbiography}[{\includegraphics[width=1in,height=1.25in,clip,keepaspectratio]{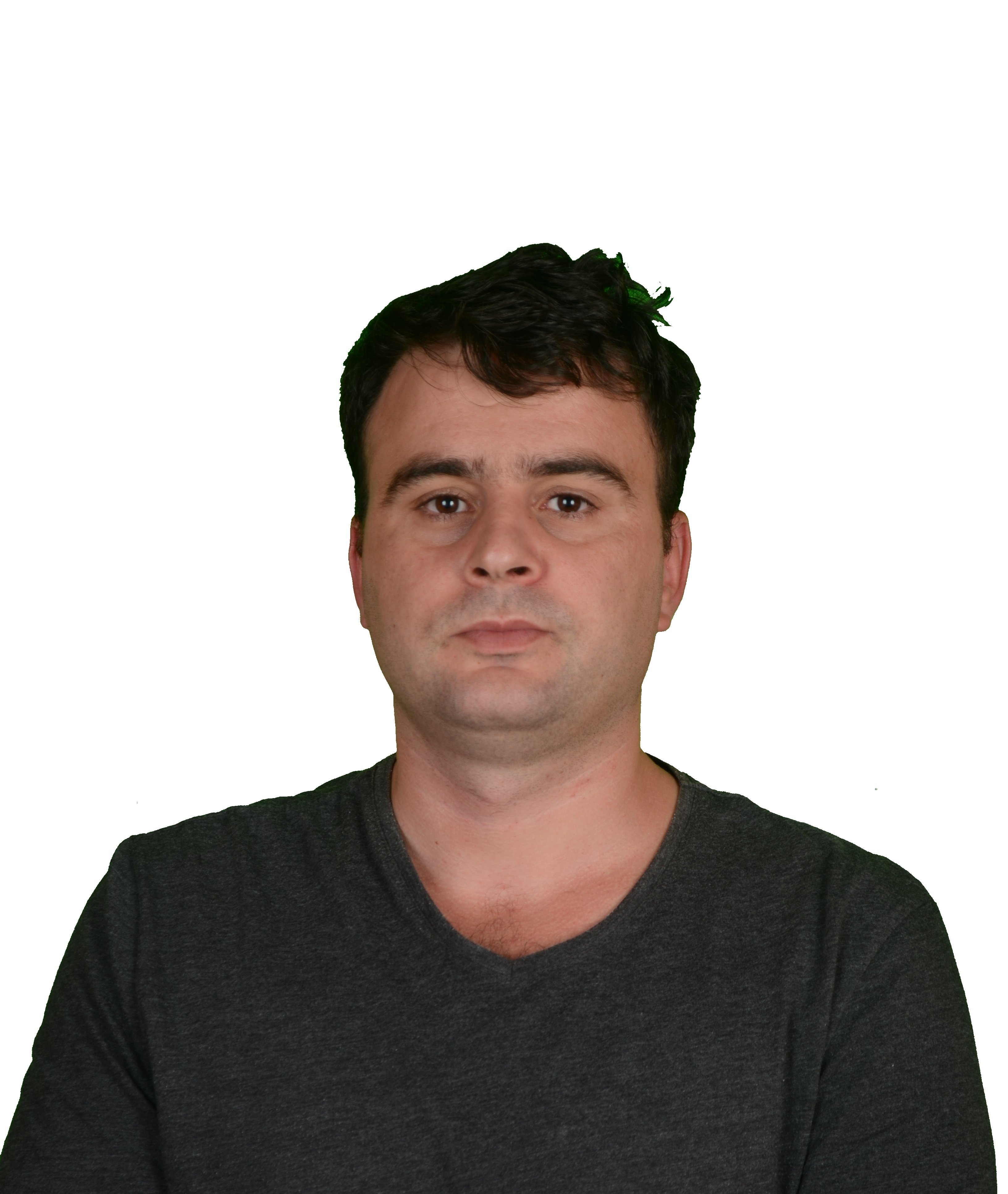}}]{Yucel Aydin}
(aydinyuc@itu.edu.tr) is a Ph.D. candidate in Cyber Security Engineering and Cryptography Program in Informatics Institute, Istanbul Technical University, Turkey. He received his MSc. Degree in the same program in 2017. His research interests are network security, cryptography and computer security.
\end{IEEEbiography}

\begin{IEEEbiography}[{\includegraphics[width=1in,height=1.25in,clip,keepaspectratio]{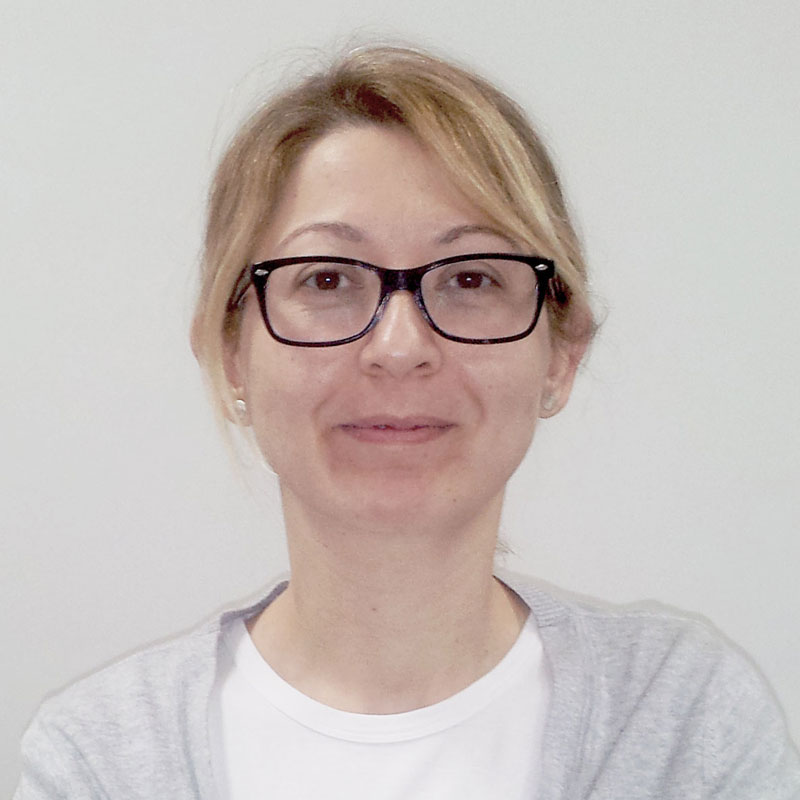}}]{Gunes Karabulut Kurt}
[StM’00, M’06, SM’15] (gkurt@itu.edu.tr) received the Ph.D. degree in electrical engineering from the University of Ottawa, Ottawa, ON, Canada, in 2006. Between 2005-2008, she was with TenXc Wireless, and Edgewater Computer Systems, in Ottawa Canada. From 2008 to 2010, she was with Turkcell R\&D Applied Research and Technology, Istanbul. Since 2010, she has been with ITU. She is an Adjunct Research Professor at Carleton University, and serving as an Associate Technical Editor of IEEE Communications Magazine.
\end{IEEEbiography}

\begin{IEEEbiography}[{\includegraphics[width=1in,height=1.25in,clip,keepaspectratio]{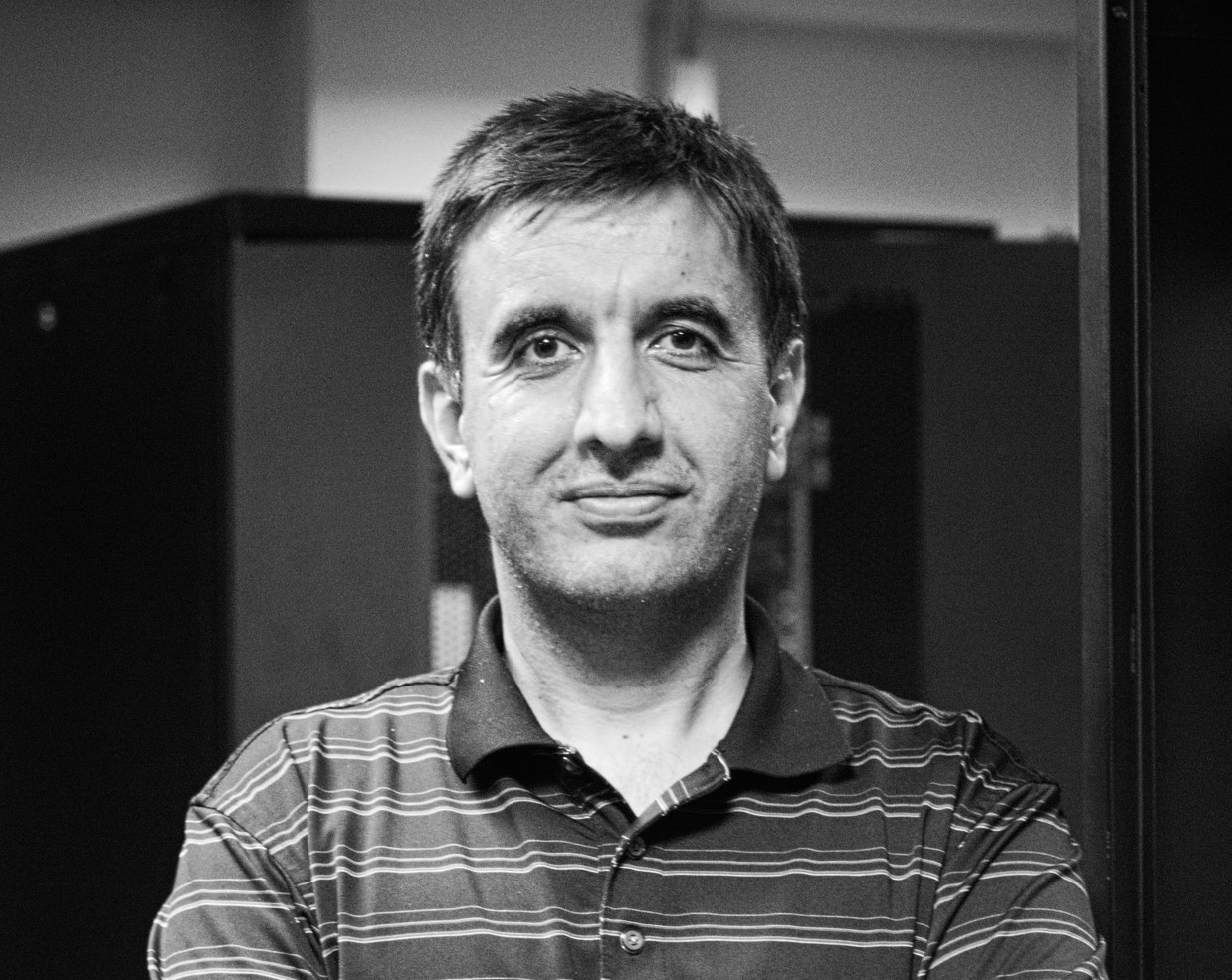}}]{Enver Ozdemir}
(ozdemiren@itu.edu.tr) received the Ph.D. degree in mathematics from the University of Maryland College Park, College Park, MD, USA, in 2009. He is currently an Associate Professor with Informatics Institute, Istanbul Technical University, Istanbul, Turkey. He was a member of the Coding Theory and Cryptography Research Group, Nanyang Technological University, Singapore from 2010 to 2014. His research interests include cryptography, computational number theory, and network security.
\end{IEEEbiography}

\begin{IEEEbiography}[{\includegraphics[width=1in,height=1.25in,clip,keepaspectratio]{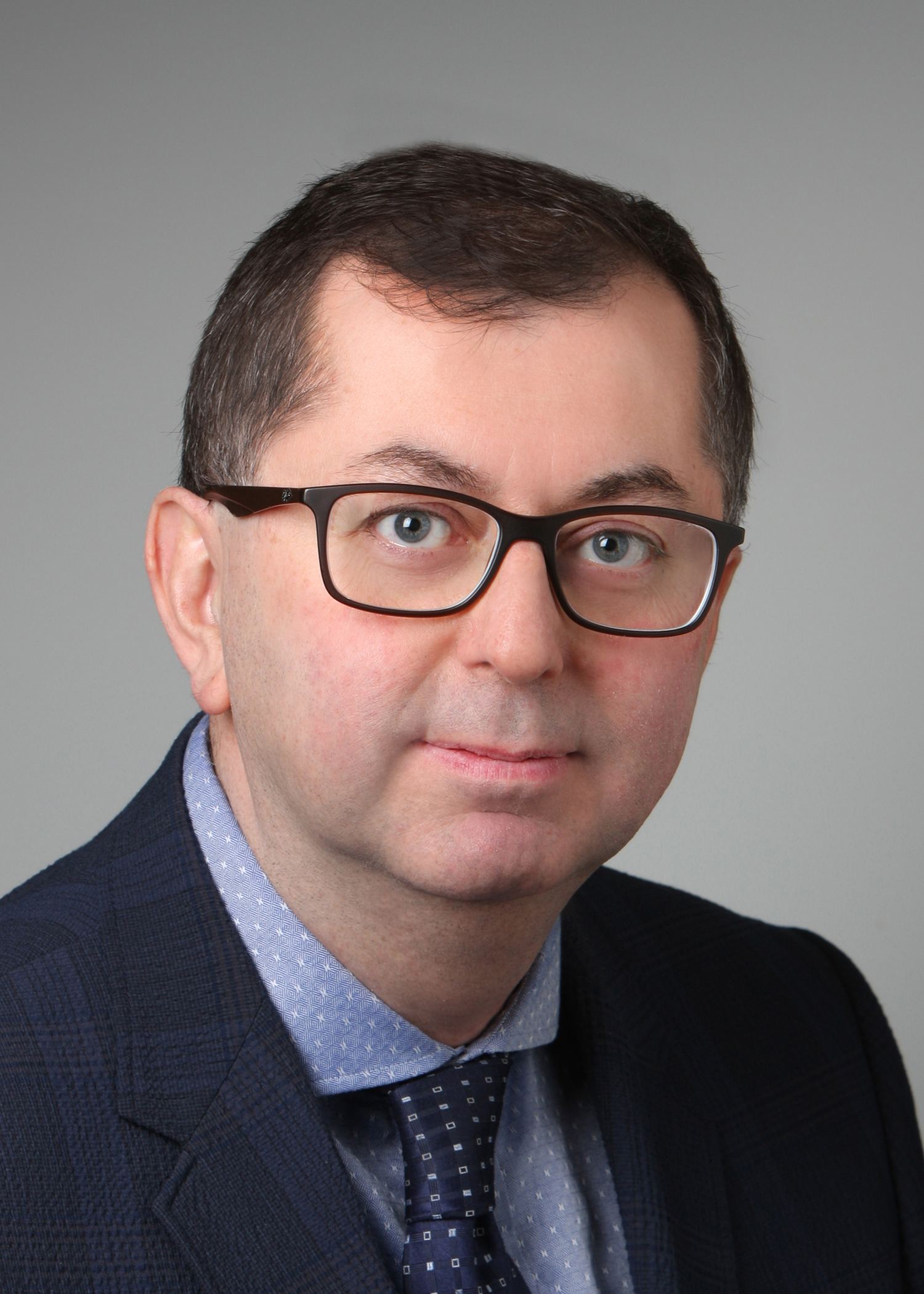}}]{Halim Yanikomeroglu}
[F] (halim@sce.carleton.ca) is a full professor in the Department of Systems and Computer Engineering at Carleton University, Ottawa, Canada. His research interests cover many aspects of 5G/5G+ wireless networks. His collaborative research with industry has resulted in 37 granted patents. He is a Fellow of the Engineering Institute of Canada and the Canadian Academy of Engineering, and he is a Distinguished Speaker for IEEE Communications Society and IEEE Vehicular Technology Society.
\end{IEEEbiography}

\end{document}